\RequirePackage{fix-cm}
\documentclass[smallextended]{svjour3}       
\smartqed  
\usepackage{graphicx}

\usepackage{amsmath,amsthm,amsfonts}
\usepackage{amssymb,mathrsfs}
\usepackage{hyperref}
\usepackage{float}
\usepackage{relsize}
\usepackage{siunitx}
\usepackage{geometry}
\usepackage{dcolumn}
\usepackage{bm}
\newcommand{\lambdabar}{{\mkern0.75mu\mathchar '26\mkern -9.75mu\lambda}}
\begin{document}

\title{Gravitoelectromagnetism in metric $f(R)$ and Brans--Dicke theories with a potential}


\author{A. Dass        \and
        S. Liberati 
}


\institute{A. Dass \at
               Dipartimento di Fisica, Universit\`a degli Studi di Trento, Via Sommarive, 14, 38123 Povo, Trento TN, Italy and SISSA, Italy \\
              \email{abhinandan.dass@alumni.unitn.it}           
           \and
           S. Liberati \at
              SISSA, Via Bonomea, 265, 34136 Trieste TS, Italy and INFN, sezione di Trieste, Italy.\\
              \email{liberati@sissa.it}
}

\date{Received: date / Accepted: date}

\maketitle

\begin{abstract}
A Gravitoelectromagnetism formalism in the context of metric $f(R)$ theory is presented and the analogue Lorentz force law is derived. Some interesting results such as the dependence of the deviation from General Relativity on the absolute value of the scalar potential are found, it is also found that the $f(R)$ effects are only relevant at a shorter distance or when the distance is much less than the compton wavelength, and that the effects are attractive in nature. An investigation of gravitational time delay in the context of metric $f(R)$ is also presented showing that the Ricci scalar alone is responsible for the time delay effect which seems to suggest that the extra scalar degree of freedom associated to $f(R)$ does not provide any modification. Also, to generalise our results, the Lorentz force law and gravitational time delay in the case of Brans--Dicke theories with a potential are derived; it is shown that the results are consistent with those obtained in the case of metric $f(R)$ and General Relativity in the appropriate limits.
\end{abstract}

\section{Introduction: Gravitolelectromagnetism}
\label{intro}
The parallelism between Gravitation and Electromagnetism can be traced back to the eighteenth century when Charles--Augustin de Coulomb forumlated the empirical inverse square Coloumb's law, $\vec{F_e}=K \dfrac{q_1q_2}{|\vec{r}|^2}\vec{\hat{r}}$, \cite{griffiths2005introduction} describing the force between two charges $q_1$ and $q_2$ at a distance $r$ where $K$ is the Coloumb's constant. The form of the equation had a stark similarity with the Gravitational force law \cite{newton1972principia}, $\vec{F_G}=-G\dfrac{m_1m_2}{|\vec{r}|^2}\vec{\hat{{r}}}$ where $G$ is the Gravitational constant and $m_1$ and $m_2$ are two masses. The analogy between gravity and electrodynamics also led to a postulate by Holzm\"uller \cite{Holzmuller} and Tisserand \cite{tisserand1872mouvement,tisserand1890mouvement} that the force of gravitation exerted by the Sun on the planets has an extra magnetic force leading to the precession of orbits and hence could account for the discrepancy found by Newton in the precession of the orbit of Mercury. While a full explanation of Mercury perihelion precession was achieved only with the advent of General Relativity (GR), it is still true that the analogy with electromagnetism has been proved useful over time, at least in some regimes. Indeed, it can be further demonstrated that gravitation contains a gravitomagnetic field due to the mass current \cite{thirring1918effect,thirring1918wirkung,lense1918influence,mashhoon2003gravitational,ciufolini1995gravitation}. In fact, to bring Newton's gravity and Lorentz invariance under the same roof, the concept of gravitomagnetic field is a must \cite{mashhoon1984gravitational}. It is known that Einstein's Relativity predicts a gravitomagnetic field due to the proper rotation of the Sun which influences the planetary orbits. This was first studied by de Sitter \cite{de1916w} and then in a more consistent general form by Lense and Thirring. Ciufolini offered some insights into the gravitomagnetic field of Earth by studying the motion of laser-ranged satellites LAGEOS and LAGEOS II \cite{ciufolini20001995}. Mashhoon et al studied the stress-energy tensor of Gravitoelectromagnetism (GEM) in \cite{mashhoon2001gravitoelectromagnetism,mashhoon1997gravitational} and the gravitational Larmor theorem in \cite{mashhoon1993gravitational}; and a brief review of GEM in GR is given in \cite{mashhoon2003gravitoelectromagnetism}. A precession experiment to measure the Lense--Thirring precession and geodetic precession was also conducted by Gravity Probe--B \cite{everitt2011gravity}.
\section{\textbf{Gravitoelectromagnetism: Metric $f(R)$ gravity}}
\label{sec:1}
General Relativity can be extended by higher than second order field equations. Such an extension will allow the graviton propagator to fall off more quickly in the UV regime and hence improve the renormalisability properties. However, it can also cause instabilities by introducing ghost degrees of freedom \cite{woodard2007avoiding}.\\\\ In $f(R)$ theories, a generic function of the Ricci scalar is allowed instead of just a linear term. The precursor of $f(R)$ were the one where additional scalar curvature invariants were added to the action instead of just the linear Ricci scalar $R$ in the action. Here, we will consider a general function of the Ricci scalar which leads to fourth--order field equations. Such theories have improved renormalisation properties \cite{stelle1977renormalization} without ghost and could also help explaining an inflationary phase \cite{starobinsky1980new}. Finally, it is worth stressing that a particular $f(R)$ Lagrangian was derived as the effective classical Lagrangian which would lead to the Friedmann equations of Loop Quantum Cosmology, both within a metric and a Palatini ansatz \cite{sotiriou2009covariant,olmo2009covariant}.\\\\
For the present analysis, a linearised metric $f(R)$ theory wherein $f(R)$ is considered to be an analytic function about $R=0$ is considered. The time-like convention of Landau-Lifschitz $(+---)$ is used. \\\\
The choice of an analytic expansion of $f(R)$ is made because of the following reasons:
\begin{itemize}
\item $\dfrac{1}{R}$ models do not seem to have the correct Newtonian limit and there is no clear evidence that they pass the solar system tests as mentioned in \cite{sotiriou2007metric}.
\item In \cite{dolgov2003can}, it was shown that $\dfrac{1}{R}$ models lead to instability in the weak gravity regime. 
\end{itemize}
The choice of a metric theory, instead of the Palatini one is made because in this case, even a simple polytropic equation of state would lead to a curvature singularity for a static spherically symmetric solution \cite{dolgov2003can}. Moreover, the Palatini theory can be shown to be classically equivalent to a singular generalized Brans--Dicke theory as we shall see later. \\\\
Following the procedure of linearisation in \cite{berry2011linearized}, an analytic $f(R)$ can be expanded around $R=0$ as 
\begin{equation}
f(R)=a_0+a_1R+\frac{a_2}{2!}R^2+\frac{a_3}{3!}R^3+\hdots
\end{equation}
The dimension of $f(R)$ has to be the same as $R$, and hence $[a_{n}]=[R]^{(1-n)}$. Also, the requirement of correct GR limit implies that $a_{1}=1$, any rescaling will be absorbed in the definition of $G$.\\\\The vacuum field equations are given by
\begin{equation}
f^\prime {R_{\mu \nu}} - \nabla_\mu \nabla_\nu f^\prime + {g_{\mu \nu}}\square f^\prime - \frac{f}{2}{g_{\mu \nu}} = 0.
\end{equation}
Taking the trace gives us
\begin{equation}\label{1.105}
f^\prime R + 3\square f^\prime -2f = 0.
\end{equation}
Note that for a flat spacetime, $R=0$, this equation gives \cite{capozziello2007newtonian}
\begin{equation}\label{1.106}
a_0=0,
\end{equation}
which is equivalent to saying that there is no cosmological constant. We define the following in analogy with the Einstein tensor of GR
\begin{equation}\label{1.107}
\mathcal{G}_{\mu \nu} \equiv f^\prime R_{\mu \nu} - \nabla_\mu \nabla_\nu f^\prime+ g_{\mu \nu}\square f^\prime - \frac{f}{2}g_{\mu \nu},
\end{equation}
such that in vacuum, we have
\begin{equation}
\mathcal{G}_{\mu \nu}=0.
\end{equation}
Similar to standard analysis done in GR, we will consider the perturbed metric about a Minkowski background, $g_{\mu\nu}=\eta_{\mu\nu}+h_{\mu\nu}$. The linearised connection is given by
\begin{equation}
\Gamma^{(1)^\rho}_{\mu \nu}=\frac{1}{2}\eta^{\rho \lambda}(\partial_\mu h_{\lambda \nu} + \partial_\nu h_{\lambda \mu} - \partial_\lambda h_{\mu \nu}),
\end{equation}
and the linearised Riemann tensor is given by
\begin{equation}
R^{(1)^\lambda}_{\mu \nu \rho}=\frac{1}{2}(\partial_\mu \partial_\nu h^\lambda_\rho + \partial^\lambda \partial_\rho h_{\mu \nu} - \partial_\mu \partial_\rho h^\lambda_\nu - \partial^\lambda \partial_\nu h_{\mu \rho}),
\end{equation}
where as usual we have raised and lowered the indices using the flat metric.\\\\
The linearised Ricci tensor can be obtained by the usual self contraction of the above given linearised Riemann tensor
\begin{equation}\label{1.111}
R_{\mu\nu}^{(1)}=\frac{1}{2}(\partial_\mu \partial_\rho h^\rho_\nu+\partial_\nu \partial_\rho h^\rho_\mu-\partial_\mu \partial_\nu h-\Box h_{\mu\nu}),
\end{equation}
and a further contraction with the flat geometry gives the Ricci scalar
\begin{equation}\label{1.112}
R^{(1)}=\partial_\mu \partial_\rho h^{\rho \mu} - \square h.
\end{equation}
To first order in $h_{\mu\nu}$, we can express $f(R)$ as a Maclaurin series
\begin{equation}\label{377}
f(R)=a_0+R^{(1)},
\end{equation}
\begin{equation}\label{378}
f^\prime=1+a_2R^{(1)},
\end{equation}
We are perturbing around a Minkowski background where the Ricci scalar vanishes, we make use of equation \eqref{1.106} to set $a_0=0$ and inserting the resulting equations in \eqref{1.107}, we get
\begin{equation}\label{1.115}
\mathcal{G}^{(1)}_{\mu \nu} = R^{(1)}_{\mu\nu}-\partial_\mu \partial_\nu(a_2R^{(1)}) + \eta_{\mu \nu}\square(a_2R^{(1)})- \frac{R^{(1)}}{2}\eta_{\mu \nu}.
\end{equation}
While from the linearised trace equation, \eqref{1.105}, we obtain
\begin{equation}\label{1.116}
\mathcal{G}^{(1)} = 3\square(a_2R^{(1)})-R^{(1)}.
\end{equation}
where $\mathcal{G}^{(1)}=\eta^{\mu\nu}\mathcal{G}_{\mu\nu}^{(1)}$. This is the massive inhomogeneous Klein--Gordon equation. Setting $\mathcal{G}=0$ for a vacuum solution in $f(R)$ at all orders of perturbations, the standard Klein--Gordon equation is recovered
\begin{equation}\label{576}
\square R^{(1)} + \Upsilon^2 R^{(1)} = 0.
\end{equation}
Where we have defined a reciprocal length as 
\begin{equation}\label{75}
\Upsilon^2 = -\frac{1}{3a_2}.
\end{equation}
If we want a physically meaningful solution, we require $\Upsilon^2>0$ and hence we constrain $f(R)$ such that $a_2<0$ \cite{schmidt1986h,teyssandier1990new,olmo2005gravity,corda2008massive}. From $\Upsilon$, we define a reduced Compton wavelength associated with the scalar mode. 
\begin{equation}
\lambdabar=\frac{1}{\Upsilon}.
\end{equation}
In order to look for wave solutions, we need to express the linearised Einstein tensor and its trace in terms of the perturbation of the background geometry, $h_{\mu\nu}$ and $h$. As such, a quantity $\bar{h}_{\mu\nu}$ is required that will satisfy a wave equation and is related to $h_{\mu\nu}$ as
\begin{equation}
\bar{h}_{\mu \nu} = h_{\mu \nu} + A_{\mu \nu}.
\end{equation}
In GR, one normally uses the trace-reversed form, where $A_{\mu\nu}=-\dfrac{h}{2}\eta_{\mu\nu}$. It is evident that this will not be sufficient in this case for a wave solution but we shall look for a solution on similar lines by introducing the ansatz
\begin{equation}\label{1.120}
\bar{h}_{\mu \nu} = h_{\mu \nu} - \frac{h}{2}\eta_{\mu \nu} + B_{\mu \nu}.
\end{equation}
Where $B_{\mu\nu}$ is a symmetric rank-2 tensor. The only rank-two tensors in our theory are $h_{\mu\nu}$, $\eta_{\mu\nu}$, $R_{\mu\nu}^{(1)}$, and $\partial_{\mu}\partial_{\nu}$. Let us notice that $B_{\mu\nu}$ needs to be first order in
$h$, and depend on $f(R)$. This can be easily achieved by the following ansatz \cite{berry2011linearized}
\begin{equation}\label{1.121}
\bar{h}_{\mu \nu} = h_{\mu \nu} + \Big(a_2bR^{(1)}-\frac{h}{2}\Big)\eta_{\mu \nu},
\end{equation}
where $a_2$ has been introduced for dimensional consistency and $b$ is a dimensionless number. The contraction with the flat metric gives
\begin{equation}
\bar{h}=4a_2bR^{(1)}-h.
\end{equation}
We can hence eliminate $h$ in \eqref{1.121} to give
\begin{equation}\label{76}
h_{\mu \nu} = \bar{h}_{\mu \nu} + \Big(a_2bR^{(1)}-\frac{\bar{h}}{2}\Big)\eta_{\mu \nu}.
\end{equation}
In analogy with GR, we have the freedom to perform a gauge transformation \cite{weinberg2014gravitation} given that the field equations are gauge invariant, the Lagrangian being a function of gauge invariant Ricci scalar. Following the usual treatment in GR, we assume a de Donder gauge
\begin{equation}
\nabla^\mu \bar{h}_{\mu \nu} = 0,
\end{equation}
which in flat spacetime gives
\begin{equation}
\partial^\mu \bar{h}_{\mu \nu} = 0,
\end{equation}
Now, subject to the above conditions, the Ricci tensor \eqref{1.111} becomes
\begin{equation}
R^{(1)}_{\mu \nu} = -\frac{1}{2} \Big[ 2b\partial_\mu \partial_\nu (a_2R^{(1)}) + \square \Big(\bar{h}_{\mu \nu} - \frac{\bar{h}}{2}\eta_{\mu \nu} \Big)+ \frac{b}{3}(R^{(1)} + \mathcal{G}^{(1)})\eta_{\mu \nu} \Big].
\end{equation}
Contraction of the above gives
\begin{equation}
R^{(1)} = -\frac{1}{2}\Big[2b\Box(a_2R^{(1)})-\Box \bar{h}+\frac{4b}{3}(R^{(1)}+\mathcal{G}^{(1)})\Big].
\end{equation}
If we replace $G^{(1)}$ above by equation \eqref{1.116}, we get the Ricci scalar
\begin{equation}\label{753}
R^{(1)} = -3b\Box(a_2R^{(1)})+\frac{1}{2}\Box\bar{h}.
\end{equation}
Using the above expression in \eqref{1.115} gives
\begin{equation}\label{1.128}
\mathcal{G}^{(1)}_{\mu \nu} = \frac{2-b}{6}\mathcal{G}^{(1)}\eta_{\mu \nu} - \frac{1}{2}\square\Big(\bar{h}_{\mu \nu} - \frac{\bar{h}}{2}\eta_{\mu \nu} \Big) - (b+1)\Big[\partial_\mu \partial_\nu (a_2R^{(1)}) + \frac{1}{6}R^{(1)}\eta_{\mu \nu} \Big].
\end{equation}
The ansatz given in \eqref{1.121} is a convenient redefintion of the gravitational perturbations so as to simplify the resolution of the wave equation. In our ansatz \eqref{1.121}, we have a separate dimensional coefficient $a_2$ and an arbitrary dimensionless rescaling by a number $b$. Here, \eqref{1.128} shows that simplest form of the wave solution for the perturbed Einstein tensor is obtained by picking up the ansatz \eqref{1.121} with $b=-1$. A such, without loss of generality, we can fix $b=-1$, so that the last term above vanishes and equations \eqref{1.121} and \eqref{76} becomes \cite{corda2008massive,capozziello2008massive}
\begin{equation}
\bar{h}_{\mu \nu} = h_{\mu \nu} - \Big(a_2R^{(1)}+\frac{h}{2}\Big)\eta_{\mu \nu}.
\end{equation}
\begin{equation}
h_{\mu \nu} = \bar{h}_{\mu \nu} - \Big(a_2R^{(1)}+\frac{\bar{h}}{2}\Big)\eta_{\mu \nu}.
\end{equation}
From \eqref{753}, the Ricci scalar becomes
\begin{equation}
R^{(1)} = 3\square(a_2R^{(1)}) + \frac{1}{2}\square\bar{h}.
\end{equation}
To be consistent with \eqref{1.116}, we need
\begin{equation}\label{659}
-\frac{1}{2}\square\bar{h}=\mathcal{G}^{(1)}.
\end{equation}
Inserting the above in \eqref{1.128} along with $b=-1$, we get
\begin{equation}\label{660}
-\frac{1}{2}\square\bar{h}_{\mu \nu}=\mathcal{G}^{(1)}_{\mu \nu}.
\end{equation}
If $a_2$ is sufficiently small that it can be neglected, then equations \eqref{659} and \eqref{660} drops down to that of GR.\\\\
We now add a source term $T_{\mu\nu}$, the linearised equations are found at the first order in perturbation
\begin{equation}\label{499}
-{\frac{1}{2}}\Box \bar {h}=G^{(1)}={\frac{8 \pi G}{c^4}}T,
\end{equation}
\begin{equation}\label{456}
-{\frac{1}{2}}\Box \bar {h}_{\mu \nu}=G^{(1)}_{\mu\nu}={\frac{8 \pi G}{c^4}}T_{\mu \nu},
\end{equation}
which can be rewritten as
\begin{equation}\label{35}
\Box \bar {h}_{\mu \nu}=-{\frac{16 \pi G}{C^4}}T_{\mu \nu},
\end{equation}
which is the wave equation for the tensor mode.
Also, for the scalar mode, using \eqref{499} in \eqref{1.116} and remembering \eqref{75}, the following equation has to be solved
\begin{equation}\label{34}
\Box  R^{(1)} + \Upsilon ^2 R^{(1)}={\frac{8 \pi G}{c^4}}\Upsilon^2 T.
\end{equation}
To solve the above two equations, \eqref{35} and \eqref{34} with the source, let us introduce the Green function
\begin{equation}
(\Box +\Upsilon^2)\mathscr{G}_\Upsilon (x,x')=\delta(x-x')
\end{equation}
where $\Box$ acts on $x$, and $\mathscr{G}_\Upsilon$ is given by
\begin{equation}
\mathscr{G}_\Upsilon (x,x')=\frac{1}{(2\pi)^4}\int d^4p\frac{\exp[-ip\cdot(x-x')]}{\Upsilon^2-p^2}.
\end{equation}
The above can be solved by the method of contour integration to give
\begin{equation}\label{2.81}
\mathscr{G}_\Upsilon (x,x')=
\begin{cases}
\mathlarger{\int}{\dfrac{d\omega}{2\pi}}\exp[-i\omega(t-t')]{\frac{1}{4\pi r}}\exp[i(\omega^2 - \Upsilon^2)^{{\frac{1}{2}}}r], \quad \omega^2>\Upsilon^2\\
\mathlarger{\int}{\dfrac{d\omega}{2\pi}}\exp[-i\omega(t-t')]{\frac{1}{4\pi r}}\exp[-(\Upsilon^2-\omega^2 )^{{\frac{1}{2}}}r],\quad \omega^2<\Upsilon^2\\
\end{cases}
\end{equation}
where we have, $t=x^{0}$, $t'=x'^{0}$ and $r=|x-x'|$.\\\\ The tensor equation \eqref{35} does not have an
associated mass (the graviton is stil massless). So, in this case, the relevant Green function is the one for $\Upsilon=0$
\begin{equation*}
\mathscr{G}_{0} (x,x')=\frac{\delta(t-t'-r)}{4\pi r},
\end{equation*}
which is the usual retarded time Green function. Using it to solve \eqref{35}, one gets
\begin{equation*}
\bar{h}_{\mu\nu}=-{\frac{16\pi G}{c^4}}\int d^4x' \mathscr{G}_{0} (x,x') T_{\mu\nu}(x'),
\end{equation*}
\begin{equation}\label{2.79}
 \bar {h}_{\mu \nu}=-{\frac{4G}{c^4}}\int d^3x' {\frac{T_{\mu \nu}(t-r,x')}{r}}.
\end{equation}
The scalar mode equation \eqref{34} is instead solved as
\begin{equation}\label{692}
R^{(1)}(x)=-8\pi G\Upsilon ^2\int d^4x' \mathscr{G}_\Upsilon (x,x') T(x').
\end{equation}
Let's go to the Newtonian limit and consider a stationary, point source mass distribution. We have
\begin{align}\label{2.82}
T_{00} &=\rho c^2, & T_{0i} &=-cj_i, & \rho &=M\delta (x'), &T_{ij} \approx 0
\end{align}
Where $\vec{j}=\rho\vec{v}$ is the mass current.
And 
\begin{align*}
|T_{00}| & \gg |T_{0i}|, & |T_{00}|\gg|T_{ij}|.
\end{align*}
So we get,
\begin{equation}\label{2.83}
 \bar {h}_{00}=-{\frac{4GM}{r c^2}}.
\end{equation}
 In order to find $R^{(1)}$, let's define in the equation \eqref{2.81}
\begin{equation}\label{678}
f(r,\omega)=
\begin{cases}
\exp[i(\omega^2 - \Upsilon^2)^{{\frac{1}{2}}}r],\quad \omega^2>\Upsilon^2\\
\exp[-(\Upsilon^2-\omega^2 )^{{\frac{1}{2}}}r],\quad \omega^2<\Upsilon^2\\
\end{cases}
\end{equation}
Then from equation \eqref{692}, we get
\begin{align}\label{2.85}
R^{(1)}(x) &=-8\pi G\Upsilon ^2\int d^4x' \mathscr{G}_\Upsilon (x,x') M\delta^{3}(x')\nonumber \\
&=-8\pi G\Upsilon ^2\int dt'\int {\frac{d\omega}{2\pi}}\exp[-i\omega(t-t')]M{\frac{1}{4\pi r}}f(r,\omega)\\
&=-8\pi G\Upsilon ^2 M{\frac{1}{4\pi r}}f(r,0)\nonumber \\
&=-2 G\Upsilon^2 M{\frac{\exp(-\Upsilon r)}{r}}\nonumber .
\end{align}
Using $\bar {h}=\bar {h}_{00}$, \eqref{75}, \eqref{2.83} and  \eqref{2.85} in \eqref{76}, we get
\begin{equation}\label{339}
h_{00}=-{\frac{2GM}{r}}\left[1 + \frac {\exp(-\Upsilon r)}{3}\right].
\end{equation}
We can also prove in a similar manner, that the purely space component has the following form
\begin{equation}\label{1056}
h_{ij}=-{\frac{2GM}{r}}\left[1 - \frac {\exp(-\Upsilon r)}{3}\right]\delta_{ij}.
\end{equation}
Along with this, if we extend the result and consider a slowly rotating source with angular momentun $J$, then we have an additional term $\bar {h}^{0i}=h^{0i}$ \cite{hobson2006general}.\\\\
Then, using \eqref{2.79} along with \eqref{2.82}, we get
\begin{equation}\label{2.91}
\bar {h}_{\mu \nu}=-{\frac{4G}{c^4}}\int d^3x' {\frac{T_{\mu \nu}(x')}{r}},
\end{equation}
\begin{equation}\label{2.92}
\bar {h}_{00}=-{\frac{4G}{c^4}}\int d^3x' {\frac{\rho c^2}{r}}.
\end{equation}
We can now define as usual $\bar {h}_{00}$ as
\begin{equation}\label{2.93}
\bar {h}_{00}\equiv -\frac{4\Phi}{c^2},
\end{equation}
where $\Phi$ is the scalar potential. Different authors have used different definitions for this potential. Some have taken it with an overall $+$ sign like in \cite{hobson2006general} while \cite{padmanabhan2010gravitation} uses the definition with a $-$ sign as in \eqref{2.93}.\\\\
Comparing \eqref{2.92} and \eqref{2.93}, we get
\begin{equation}\label{2.94}
\Phi=G\int d^3x'\frac{\rho}{r}.
\end{equation}
Applying the Laplacian to \eqref{2.94} and since $\vec{\nabla}^2 \dfrac{1}{r}=-4\pi\delta(r)$ \cite{redzic2013laplacian}, we get the Poisson equation
\begin{equation}\label{2.95}
\vec{\nabla}^2 \Phi=-4\pi G\rho.
\end{equation}
Now, define $\bar {h}_{0i}$ as \cite{hobson2006general}
\begin{equation}\label{2.96}
\bar {h}_{0i}=h_{0i}\equiv\frac{2A_i}{c^2},
\end{equation}
where $\vec{A}$ is the vector potential.
From \eqref{2.91} and \eqref{2.96}
\begin{equation}
\bar {h}_{0i}=-{\frac{4G}{c^4}}\int d^3x' {\frac{-c j_i}{r}}=\frac{2A_i}{c^2},
\end{equation}
which after a trivial calculation gives
\begin{equation}
A_i=\frac{2G}{c}\int d^3x'\frac{j_i}{r},
\end{equation}
applying the Laplacian to the above expression, it yields
\begin{equation}\label{2.99}
\vec{\nabla}^2 A_i=-\frac{8\pi G}{c}j_i.
\end{equation}
Now in analogy with Electromagnetism, we define the gravitoelectric field $\vec{E}$
\begin{equation}\label{2.100}
\vec{E}\equiv -\vec{\nabla} \Phi - \frac{1}{2c}\frac{\partial \vec{A}}{\partial t},
\end{equation}
and the gravitomagnetic field $\vec{B}$
\begin{equation}\label{2.101}
\vec{B}\equiv \vec{\nabla} \times \vec{A}.
\end{equation}
Let us use the gauge $\partial_{\mu} \bar {h}^{\mu \nu}=0$, so that \eqref{2.93} and \eqref{2.96} give
\begin{equation}\label{2.102}
\frac{1}{c}\frac{\partial \Phi}{\partial t} + \frac{1}{2}\vec{\nabla}\cdot \vec{A}=0.
\end{equation}
We observe that the above equation is similar to one in the Lorentz gauge in EM and is consistent with the corresponding results in GR and scalar-tensor theories \cite{bezerra2005some}.\\\\
Now, taking the divergence of the gravitoelectric field \eqref{2.100}, one obtains
\begin{equation}\label{2.103}
\vec{\nabla}\cdot \vec{E}=-\vec{\nabla}^2 \Phi - \frac{1}{2c}\frac{\partial (\vec{\nabla}\cdot \vec{A})}{\partial t},
\end{equation}
and using \eqref{2.95} in \eqref{2.103} along with the gauge condition \eqref{2.102}, one gets
\begin{equation}
\vec{\nabla}\cdot \vec{E}=4\pi G \rho.
\end{equation}
Taking the curl of \eqref{2.101}, yields
\begin{equation}
\vec{\nabla} \times \vec{B}=\vec{\nabla} \times (\vec{\nabla} \times \vec{A})=\vec{\nabla} (\vec{\nabla} \cdot \vec{A}) - \vec{\nabla}^2 \vec{A},
\end{equation}
and introducing \eqref{2.102} and \eqref{2.99} in the above expression, we finally obtain
\begin{equation}
\vec{\nabla} \times \vec{B}=\frac{2}{c}\frac{\partial E}{\partial t} + \frac{8\pi G}{c} \vec{j}.
\end{equation}
This ends the formalism of the analogue Maxwell equations in the context of metric $f(R)$. Now, to make the theory complete, we look for an analogue of the Lorentz Force law.
\subsection{\textbf{Metric $f(R)$ Gravitoelectromagnetism force law}}
\label{sec:2}
In order to derive the Lorentz force law in metric $f(R)$ theories we recall the expression of the metric perturbation for the purely time and space component, i.e, \eqref{339} and \eqref{1056}
\begin{equation}
h_{00}=-{\frac{2GM}{r}}\left[1 + \frac {\exp(-\Upsilon r)}{3}\right],
\end{equation}
\begin{equation}
h_{ij}=-{\frac{2GM}{r}}\left[1 - \frac {\exp(-\Upsilon r)}{3}\right]\delta_{ij}.
\end{equation}
Taking also into account expression \eqref{2.96}, we then get for the line element
\begin{equation}
ds^2= c^2\left\{{1-\frac{2\Phi}{c^2} \left[1+ \frac {\exp(-\Upsilon r)}{3}\right]}\right\} dt^2 + \frac{4}{c}(\vec{A} \cdot d\vec{r})dt- \left\{{1+\frac{2\Phi}{c^2} \left[1- \frac {\exp(-\Upsilon r)}{3}\right]}\right\} dr^2,
\end{equation}
where $\Phi=\dfrac{GM}{r}$ is the scalar potential.\\\\
Let's now define the following symbols 
\begin{equation}\label{2.110}
\alpha \equiv \Phi\left[1+ \frac {\exp(-\Upsilon r)}{3}\right],
\end{equation}
\begin{equation}\label{2.111}
\beta \equiv \Phi\left[1- \frac {\exp(-\Upsilon r)}{3}\right],
\end{equation}
so to write the line element as
\begin{align}\label{2.115}
ds^2 &=c^2\left(1-\frac{2\alpha}{c^2}\right)dt^2 + \frac{4}{c}\left(\vec{A} \cdot d\vec{r}\right)dt - \left(1+\frac{2\beta}{c^2}\right)dr^2 \nonumber\\
&=dt^2\left[c^2-2\alpha+\frac{4}{c}\vec{A} \cdot \vec{v}-v^2-2\frac{\beta}{c^2}v^2\right]\nonumber\\
&=dt^2\left\{c^2\left[1-\frac{v^2}{c^2}-\frac{2\alpha}{c^2}+\frac{4}{c^3}\vec{A} \cdot \vec{v} - 2\frac{\beta}{c^4}v^2\right]\right\}\nonumber\\
&=dt^2\left\{c^2\left[\frac{1}{\gamma^2}-\frac{2\alpha}{c^2}+\frac{4}{c^3}\vec{A} \cdot \vec{v} - 2\frac{\beta}{c^4}v^2\right]\right\}\nonumber\\
&=dt^2\frac{c^2}{\gamma^2}\left[1-\frac{2\alpha}{c^2}\gamma^2+\frac{4}{c^3}\vec{A} \cdot \vec{v}\gamma^2 - 2\frac{\beta}{c^4}v^2\gamma^2\right],
\end{align}
where $\gamma=\dfrac{1}{\sqrt{1-\frac{v^2}{c^2}}}$ is the Lorentz factor and $\vec{v}=\dfrac{d\vec{r}}{dt}$. 
Taking the square root on both sides of equation \eqref{2.115} and using the binomial expansion upto first order in $\alpha$ and $\beta$, we get
\begin{align}
\frac{ds}{dt} &=\frac{c}{\gamma}\left[1-\frac{\alpha}{c^2}\gamma^2+\frac{2}{c^3}\vec{A}\cdot \vec{v}\gamma^2 - \frac{\beta}{c^4}v^2\gamma^2\right]\nonumber\\
&=\frac{c}{\gamma}-\frac{\alpha}{c}\gamma+\frac{2}{c^2}\vec{A}\cdot \vec{v}\gamma - \frac{\beta}{c^3}v^2\gamma.
\end{align}
The Lagrangian is given by $\mathscr {L} = -mc\dfrac{ds}{dt}$, where $m$ is the mass of the test particle, which gives
\begin{equation}\label{2.119}
\mathscr {L}= -mc^2\left(1-\frac{v^2}{c^2}\right)^{\frac{1}{2}}+m\gamma\alpha+m\gamma\frac{v^2}{c^2}\beta - \frac{2m}{c}\gamma\vec{A}\cdot\vec{v}.
\end{equation}
Now, to find the equation of motion, we have to use the Euler--Lagrange equation, $\dfrac{d}{dt}\left(\dfrac{\partial \mathscr {L}}{\partial \vec{v}}\right)=\dfrac{\partial \mathscr {L}}{\partial \vec{x}}$,
but before that we expand the Lagrangian expression \eqref{2.119} upto first order in $\dfrac{v^2}{c^2}$ except for the first term. So, we get
\begin{equation}
\mathscr {L}=-mc^2\left(1-\frac{v^2}{c^2}\right)^{\frac{1}{2}}+m\alpha+m\frac{\alpha}{2} \frac{v^2}{c^2} +m\frac{v^2}{c^2}\beta -\frac{2}{c}m\vec{A}\cdot\vec{v}.
\end{equation}
The equation of motion can now be derived by just calculating the Euler--Lagrange equation from this Lagrangian. First of all, we get
\begin{equation}\label{2.121}
\frac{\partial\mathscr {L}}{\partial \vec{v}}=-\gamma m \vec{v}+m\alpha \frac{\vec{v}}{c^2}+2m\beta\frac{\vec{v}}{c^2}-\frac{2m}{c}\vec{A}.
\end{equation}
In this expression, we can safely ignore the second and third term being suppressed by $c^2$. Let us take the total derivative with respect to time only of the first and last term in the r.h.s of \eqref{2.121}. We get
\begin{equation}\label{2.122}
\frac{d}{dt}\left(\frac{\partial \mathscr {L}}{\partial \vec{v}}\right)=\vec{F}-\frac{2m}{c}\frac{\partial \vec{A}}{\partial t}-\frac{2m}{c}(\vec{v}\cdot\vec{\nabla} )\vec{A}.
\end{equation}
Where $F \equiv \dfrac{d\vec{p}}{dt}$ and $\vec{p}$ is the momentum given by $\vec{p}=\gamma m\vec{v}$. While the last two terms are coming from the fact that
\begin{equation}
\frac{d\vec{A}}{dt}=\frac{\partial\vec{A}}{\partial t}+\frac{\partial \vec{A}}{\partial x_i}\frac{dx_i}{dt}.
\end{equation}
Equation \eqref{2.122} is the l.h.s of our Euler--Lagrange equation. In order to compute the r.h.s, we need to vary the Lagrangian w.r.t. space
\begin{equation}\label{2.124}
\frac{\partial\mathscr {L}}{\partial \vec{x}}=m\vec{\nabla} \alpha+\frac{m}{2}\frac{v^2}{c^2}\vec{\nabla} \alpha+m\frac{v^2}{c^2}\vec{\nabla}\beta-\frac{2m}{c}\vec{\nabla} (\vec{A} \cdot \vec{v}).
\end{equation}
We shall consistently neglect again the second and third term above since they are $\dfrac{v^2}{c^2}$ suppressed and hence negligible. Now, to evaluate the last term in the above expression, we break it down as
\begin{equation}\label{2.125}
\vec{\nabla} (\vec{A} \cdot \vec{v})= (\vec{A}\cdot \vec{\nabla})\vec{v}+(\vec{v}\cdot\vec{\nabla})\vec{A}+\vec{v}\times(\vec{\nabla}\times\vec{A}) +\vec{A}\times(\vec{\nabla}\times\vec{v}).
\end{equation}
The first term vanishes because
\begin{equation}
\left(A_j\frac{\partial}{\partial x_j}\right)\frac{\partial x_i}{\partial t}=A_j\frac{\partial \delta_{ij}}{\partial t}=0.
\end{equation}
The last term also vanishes because
\begin{equation}
\vec{\nabla} \times \vec{v}=\vec{\nabla}\times\frac{d\vec{r}}{dt}=\frac{d(\vec{\nabla}\times\vec{r})}{dt}=0.
\end{equation}
The second term in \eqref{2.125} and the third term in \eqref{2.122} cancel each other, and so by finally equating \eqref{2.122} and \eqref{2.124}, we get
\begin{equation}\label{2.128}
\vec{F}=-m\left(-\vec{\nabla}\alpha-\frac{2}{c}\frac{\partial\vec{A}}{\partial t}\right)-\frac{2m}{c}\vec{v}\times (\vec{\nabla}\times\vec{A}).
\end{equation}
Now from \eqref{2.110}, we have
\begin{equation*}
\alpha=\Phi\left[1+\frac{\exp(-\Upsilon r)}{3}\right].
\end{equation*}
So, from \eqref{2.128}, we get
\begin{equation}
\vec{F}=  -m\left[-\vec{\nabla}\left(\Phi+\Phi\frac{\exp(-\Upsilon r)}{3}\right)-\frac{2}{c}\frac{\partial\vec{A}}{\partial t}\right]  -\frac{2m}{c}\vec{v}\times\vec{B}.
\end{equation}
For the stationary case, $\dfrac{\partial \vec{A}}{\partial t}=0$, and using \eqref{2.100} the above equation reduces to
\begin{equation}
\vec{F} =-m\left[\vec{E}+\vec{E}\frac{\exp(-\Upsilon r)}{3}\right]-\frac{m}{3}\Phi\Upsilon \exp(-\Upsilon r)\vec{\nabla} r -\frac{2m}{c}\vec{v}\times\vec{B}.
\end{equation}
Working in spherical polar coordinates, $\vec{\nabla}r$ gives only the $\vec{\hat{r}}$ radial unit vector and hence, we have
\begin{equation}\label{2.131}
\vec{F}_{f(R)}=-m\vec{E}-\frac{2m}{c}\vec{v}\times\vec{B}-m\vec{E}\frac{\exp(-\Upsilon r)}{3}-\frac{m}{3}\Phi\Upsilon \exp(-\Upsilon r) \vec{\hat{r}},
\end{equation}
where we have introduced the subscript $f(R)$ to denote that it is the Lorentz force in the metric $f(R)$ theory.
\subsection{\textbf{Discussion of the results in $f(R)$}}
\begin{itemize}
\item The first two terms of expression \eqref{2.131} reproduce the GR result for the Lorentz force law while the last two terms can be seen as an imprint of the effective higher terms of $f(R)$ theory.\\
\item If $\Upsilon\rightarrow\infty$, i.e., $a_2\rightarrow 0$, which means $f(R)=a_0+a_1R$, we recover the case of GR as the last two terms in the expression for force go away. I.e., our result is consistent with GR.\\
\item If $r \ll \lambdabar_R$ (Compton wavelength) where $\lambdabar_R=\dfrac{1}{\Upsilon}$, then $f(R)$ effects are relevant. We infer that $f(R)$ corrections are effective only at a short range in the context of our theory.\\
\item $f(R)$ gives only attractive corrections and is consistent with the fact that $f(R)$ is well known to have a massless graviton plus a scalar field as propagating modes \cite{corda2008massive,capozziello2008massive,capozziello2010testing,prasia2014detection}.\\
\item The effects introduced by $f(R)$ only affects the radial direction which is also a consequence of the consideration of a  spherically symmetric solution.\\
\item The last term is peculiar as it tells us that not only the change in potential is important but the absolute value of potential also dictates the $f(R)$ effects. This is because of the new scale introduced by $\Upsilon$ which has dimensions of $length^{-1}$\\
\item If we rearrange the expression of the force as the following
\begin{equation*}
\vec{F}_{f(R)}=  -m\vec{E}\left[1+\dfrac{\exp(-\Upsilon r)}{3}\right]-\dfrac{2m}{c}\vec{v}\times\vec{B}-\dfrac{m}{3}\Phi\gamma \exp(-\Upsilon r)\vec{\hat{r}}.
\end{equation*}
We can say that $f(R)$ improves upon the gravitoelectric field at short ranges. 
\end{itemize}
\subsection{\textbf{Gravitational Time Delay}}
The gravitomagnetic time delay in the case of GR was studied by Ciufolini et al. in \cite{ciufolini2003gravitomagnetic}. Using the same method and terminologies, the spacetime metric is again perturbed around a Minkowskian background as $g_{\mu\nu}=\eta_{\mu\nu}+h_{\mu\nu}$. The propagation of electromagnetic (EM) waves in the absence of perturbing potentials, $h_{\mu\nu}(x)$, are straight lines defined by the equation
\begin{equation}\label{724}
\frac{d\vec{x}^i}{dt}=c\vec{\hat{k}}^i,
\end{equation}
where $\vec{\hat{k}}^i$ is the contant unit propagation vector of the signal.\\\\
For the evaluation of gravitational time delay to the first order in $h_{\mu\nu}$, only the null condition is required \cite{ciufolini2003gravitomagnetic}, i.e. $ds^2=0$ or
\begin{equation}\label{766}
g_{\mu\nu}dx^{\mu}dx^{\nu}=0.
\end{equation}
Here, we are investigating the consequence of the condition \eqref{766} for the propagation of a EM wave on a Minkwoski background. Hence, \eqref{766} can also be rewritten as
\begin{equation}\label{794}
c^2dt^2-|d\vec{x}|^2=h_{\mu\nu}dx^{\mu}dx^{\nu},
\end{equation}
where $|d\vec{x}|^2=\delta_{ij}dx^{i}dx^{j}$. The bending of EM waves can be neglected in the first order perturbation and we can write $h_{\mu\nu}dx^{\mu}dx^{\nu}=c^2h_{\alpha\beta}k^{\alpha}k^{\beta}dt^2$ using \eqref{724}, where $k^{\alpha}=(1,\vec{\hat{k}})$ such that $n_{\mu\nu}k^{\mu}k^{\nu}=0$. Hence \eqref{794} can be written after doing a binomial expansion and considering the first order terms as
\begin{equation}
cdt=\left(1+\frac{1}{2}h_{\alpha\beta}k^{\alpha}k^{\beta}\right)|d\vec{x}|.
\end{equation}
Now, consider a EM wave that propagates from a point $P_1:(ct_1,\vec{x_1})$ to a point $P_2:(ct_2,\vec{x_2})$ in the background inertial frame, which then gives us
\begin{equation}
t_2-t_1=\frac{1}{c}|\vec{x_2}-\vec{x_1}|+\frac{1}{2c}k^{\alpha}k^{\beta}\int_{P_1}^{P_2}h_{\alpha\beta}(x)dl,
\end{equation}
where $dl=|d\vec{x}|$ is the Euclidean length element along the straight lines that joins $P_{1}$ to $P_{2}$. Hence, the gravitational time delay $\Delta$ is given by
\begin{equation}\label{2.133}
\Delta=\frac{1}{2c}k^{\alpha}k^{\beta}\int_{P_1}^{P_2}h_{\alpha\beta}(x)dl.
\end{equation}
Now, from \eqref{76}, we have
\begin{equation*}
h_{\mu \nu}=\bar {h}_{\mu \nu} - \left(a_2 R^{(1)} + {\frac{\bar {h}}{2}}\right) \eta_{\mu \nu}, 
\end{equation*}
Which we can put in \eqref{2.133} to get the expression for time delay in terms of $\bar {h}_{\mu \nu}$. Taking into account the fact that the propagation vectors are null with respect to the flat metric, \eqref{2.133} becomes
\begin{equation}
\Delta=\frac{1}{2c}\int_{P_1}^{P_2} \bar {h}_{\mu \nu}k^{\mu}k^{\nu}dl.
\end{equation}
Now, using the definition of scalar and vector potential \eqref{2.93} and \eqref{2.96}, we have
\begin{equation}
\Delta=-\frac{2}{c^3}\int_{P_1}^{P_2}\Phi dl+\frac{2}{c^3}\int_{P_1}^{P_2}\vec{A}\cdot\vec{\hat{k}}dl.
\end{equation}
So, the Shapiro time delay is given by
\begin{equation}
\Delta_s=-\frac{2}{c^3}\int_{P_1}^{P_2}\Phi dl.
\end{equation}
And, the gravitomagnetic time delay is given by
\begin{equation}
\Delta_{GM}=\frac{2}{c^3}\int_{P_1}^{P_2}\vec{A} \cdot\vec{\hat{k}}dl.
\end{equation}
Without writing down the explicit calculations, since this is similar to the case of GR, we have the following two equations as given in \cite{ciufolini2003gravitomagnetic}.
\begin{equation}\label{1022}
\Delta_s=-\frac{2GM}{c^3}\int_{P_1}^{P_2}\frac{dl}{r},
\end{equation}
\begin{equation}\label{1023}
\Delta_{GM}=\frac{2GJ}{c^4}\left(\frac{1}{r_1}+\frac{1}{r_2}\right)\frac{\vec{\hat{J}}\cdot(\vec{\hat{r_1}}\times\vec{\hat{r_2}})}{1+\vec{\hat{r_1}}\cdot\vec{\hat{r_2}}},
\end{equation}
where $\vec{\hat{r_1}}$ and $\vec{\hat{r_2}}$ are unit vectors indicating the positions of $P_1$ and $P_2$, respectively from the centre of a sphere, and $J$ is the magnitude of the angular momentum of a slowly rotating source and $\vec{\hat{J}}$ is the unit vector along it.
\subsection{\textbf{Discussion on the $f(R)$ gravitational time delay}}
We notice that $f(R)$ has no effect on the propagation of a light beam in the weak field limit provided that an analytic exapansion of $f(R)$ is considered. This seems to suggest that the extra scalar degree of freedom associated to $f(R)$ does not modify the time delay at this order of approximation.\\\\
Also, it can be shown that in a Brans-Dicke theory, the time delay is just equal to GR multiplied by the factor $\dfrac{2\omega+3}{2\omega+4}$, where $\omega$ is the constant factor modulating the kinetic term of the scalar field\cite{barros2003gravitomagnetic}.\\\\
Similarly, in scalar-tensor theory, the time delay is equal to GR one multiplied by the factor $\dfrac{1}{1+\alpha^2(\phi_0)}$ \cite{bezerra2005some}, where $\alpha(\phi)\equiv\dfrac{\partial \ln A(\phi)}{\partial \phi}$, and $A(\phi)$ is an arbitrary function of the scalar field $\phi$, which is used to transform the physical metric, $\tilde{g}_{\mu\nu}$, of scalar-tensor theory to that of a pure rank--2 tensor metric of the Einstein frame, $g_{\mu\nu}$, by the transformation, $\tilde{g}_{\mu\nu}=A^2(\phi)g_{\mu\nu}$.
\section{Lorentz Force law for Brans-Dicke with arbitrary potential}
In previous work \cite{silva2015gravitoelectromagnetic}, the Brans--Dicke theory has been considered without a potential term which effectively means that it was not the most general case. However, metric $f(R)$ can be framed as a Brans--Dicke with a vanishing parameter, $\omega=0$, but with a non vanishing potential. We should hence expect that framing a metric $f(R)$ theory as a Brans--Dicke class will allow us to consider more general versions of $f(R)$ theories and hence do away with the \textit{a priori} assumption that $f(R)$ is an analytic function. \\\\
For that purpose, let us first see how metric $f(R)$ can be framed as a Brans--Dicke class theory.\\\\
The action of metric $f(R)$ theory is given by
\begin{equation}
\mathcal{S}=\frac{1}{2\kappa}\int d^4x\sqrt{-g}f(R)+S_M(g_{\mu\nu},\psi),
\end{equation}
where the symbols have their usual meanings and $\kappa=\dfrac{8\pi G}{c^4}$. Now, one can introduce an auxiliary field, $\chi$ and write a dynamically equivalent action \cite{sotiriou2010f}:
\begin{equation}
\mathcal{S}=\frac{1}{2\kappa}\int d^4x \sqrt{-g}(f(\chi)+f'(\chi)(R-\chi))+S_M(g_{\mu\nu},\psi).
\end{equation}
Variation with respect to $\chi$ gives
\begin{equation}
\chi=R,
\end{equation}
iff $f''(\chi)\neq 0$.\\\\
Redefining field $\chi$ by $\Phi=f'(\chi)$ and setting
\begin{equation}
V(\Phi)=\chi (\Phi)\Phi -f(\chi(\Phi)),
\end{equation}
the action becomes
\begin{equation}\label{2.147}
\mathcal{S}=\frac{1}{2\kappa}\int d^4x\sqrt{-g}(\Phi R-V(\Phi))+S_M(g_{\mu\nu},\psi).
\end{equation}
So, we immediately see that metric $f(R)$ is the action of Brans--Dicke theory with $\omega=0$ or with a vanishing kinetic term.\\\\
The Brans--Dicke action with an arbitrary potential in Jordan frame is given as 
\begin{equation}
\mathcal{S}=\int d^4x\sqrt{-g}\left\{\frac{1}{16\pi}\left[\phi R-\frac{\omega}{\phi}g^{\mu\nu}\partial_{\mu}\phi\partial_{\nu}\phi-V(\phi)\right]+ \mathcal{L}_{matter}\right\},
\end{equation}
where $\phi$ is a scalar field coupled by a dimensionless constant called the Brans--Dicke parameter, $\omega$, and $V(\phi)$ is an arbitrary potential. It should be noted that the scalar field $\phi$ here does not have the canonical dimension one and instead has dimension two like the Newton's constant.\\\\
The field equations are given by
\begin{equation}
G_{\mu\nu}=\frac{8\pi}{\phi}T_{\mu\nu}+\frac{\omega}{\phi^2}\left(\nabla_{\mu}\phi\nabla_{\nu}\phi-\frac{1}{2}g_{\mu\nu}\nabla^{\alpha}\phi\nabla_{\alpha}\phi\right)+ \frac{1}{\phi}\left(\nabla_{\mu}\phi\nabla_{\nu}\phi-g_{\mu\nu}\Box_g\phi\right)-\frac{V(\phi)}{2\phi}g_{\mu\nu},
\end{equation}
and 
\begin{equation}
\Box_g\phi=\frac{1}{2\omega+3}\left(8\pi T+\phi\frac{dV(\phi)}{d\phi}-2V(\phi)\right),
\end{equation}
where $T$ is the trace of the matter energy momentum tensor, $T_{\mu\nu}$, and $\Box_g$ is the d'Alembertian operator with respect to the full metric. Now we will see how the weak field equations looks like \cite{ozer2018linearized}. For that purpose, we consider the following expansion
\begin{align}
g_{\mu\nu} &=\eta_{\mu\nu}+h_{\mu\nu}, & \phi=\phi_0+\xi,
\end{align}
where $\phi_0$ is a constant value of the scalar field and $\xi$ is the small perturbation to the scalar field and rest of the symbols have their usual meaning. Using the above, a new tensor can be defined \cite{will1993theory}
\begin{equation}\label{2.152}
\theta_{\mu\nu}=h_{\mu\nu}-\frac{1}{2}\eta_{\mu\nu}h-\eta_{\mu\nu}\frac{\xi}{\phi_0}.
\end{equation}
and the Brans--Dicke gauge is then defined as
\begin{equation}
\nabla_{\nu}\theta^{\mu\nu}=0.
\end{equation}
The weak field equations up to second order becomes \cite{will1993theory}
\begin{equation}\label{2.154}
\Box_{\eta}\theta_{\mu\nu}=-\frac{16\pi}{\phi_0}(T_{\mu\nu}+\tau_{\mu\nu})+\frac{V_{\textrm{lin}}}{\phi_0}g_{\mu\nu},
\end{equation}
and
\begin{equation}\label{2.155}
\Box_{\eta}\xi=16\pi S.
\end{equation}
Here, $\Box_{\eta}=\eta^{\mu\nu}\partial_{\mu}\partial_{\nu}$ is d'Alembertian of the flat spacetime, other symbols have their usual meaning and the term $S$ is given by
\begin{equation}
S=\frac{1}{4\omega+6}\left[T\left(1-\frac{\theta}{2}-\frac{\xi}{\phi_0}\right)+\frac{1}{8\pi}\left(\phi\frac{dV}{d\phi}-2V\right)_{\textrm{lin}}\right]+\frac{1}{16\pi}\left(\theta^{\mu\nu}\partial_{\mu}\partial_{\nu}\xi+\frac{\partial_{\nu}\xi\partial^{\nu}\xi}{\phi_0}\right).
\end{equation}
Here, the subtext "$\textrm{lin}$" means that the terms must be properly linearised.\\\\
Note that in this derivation the relation between flat and curved spacetime d'Alembertian is used
\begin{equation}
\Box_{g}=\left(1+\frac{\theta}{2}+\frac{\xi}{\phi_0}\right)\Box_{\eta}-\theta^{\mu\nu}\partial_{\mu}\partial_{\nu}\xi-\frac{\partial_{\nu}\xi\partial^{\nu}\xi}{\phi_0}  +\mathcal{O}{(x_i^2)}.
\end{equation}
We assume that the arbitrary potential, $V$ is a well behaved function and it is Taylor expandable around a constant value, $\phi=\phi_0$, such that \cite{ozer2018linearized}
\begin{equation}
V(\phi)=V(\phi_0)+\frac{dV(\phi_0)}{d\phi}\xi+\frac{1}{2}\frac{d^2V(\phi_0)}{d\phi^2}\xi^2+...
\end{equation}
 Here, $\phi_0$ is expected to be the minimum of the potential and hence the term, $\dfrac{dV(\phi_0)}{d\phi}$, vanishes. Hence, the relevant terms in the linearised equation can be written as
\begin{align}
V(\phi)g_{\mu\nu} &\approx V(\phi_0)\eta_{\mu\nu} \nonumber,\\ 
 \left(\phi\frac{dV}{d\phi}-2V\right) & \approx\phi_0 \frac{d^2V(\phi_0)}{d\phi^2}\xi-2V(\phi_0),
\end{align}
and hence the field equations \eqref{2.154} and \eqref{2.155} become
\begin{equation}\label{2.160}
\Box_{\eta}\theta_{\mu\nu}=-\frac{16\pi}{\phi_0}T_{\mu\nu}+\frac{V_0}{\phi_0}\eta_{\mu\nu},
\end{equation}
and
\begin{equation}\label{2.161}
(\Box_{\eta}-m_s^2)\xi=\frac{8\pi T}{2\omega+3}-\frac{2V_0}{2\omega+3},
\end{equation}
where 
\begin{align}\label{2.162}
V_0 &\equiv V(\phi_0), & m_s^2 \equiv \frac{\phi_0}{2\omega+3}\frac{d^2V(\phi_0)}{d\phi^2}>0.
\end{align}
We consider a particle located at $\bar{r}=0$, where $\bar{r}^2=\bar{x}^2+\bar{y}^2+\bar{z}^2$, and $T_{\mu\nu}=M\delta(\bar{r})$.
The solution of the scalar field equation \eqref{2.161} is
\begin{equation}
\xi(\bar{r})=\frac{2M}{(2\omega+3)}\frac{\exp(-m_s \bar{r})}{\bar{r}}-\frac{V_0}{3(2\omega+3)}\bar{r}^2.
\end{equation}
The solution to \eqref{2.160} is given by
\begin{align}\label{1058}
\theta_{00} &=-\frac{4M}{\phi_0}\frac{1}{\bar{r}}+\frac{V_0}{6\phi_0}\bar{r}^2, & \theta_{xx} &=-\frac{V_0}{4\phi_0}(y^2+z^2),\\
\theta_{xx} &=-\frac{V_0}{4\phi_0}(x^2+z^2), & \theta_{xx} &=-\frac{V_0}{4\phi_0}(x^2+y^2).
\end{align}
The trace $\theta$ is given by
\begin{equation}
\theta=-\frac{4M}{\phi_0\bar{r}}+\frac{2V_0}{3\phi_0}\bar{r}^2,
\end{equation}
and from the inverse of \eqref{2.152}, we get the following
\begin{align}
h_{00} &=-\left[\frac{2M}{\phi_0\bar{r}}+\frac{V_0}{6\phi_0}\bar{r}^2+\frac{\xi}{\phi_0}\right],\\
h_{ij} &=-\left[\frac{2M}{\phi_0\bar{r}}-\frac{V_0}{12\phi_0}(\bar{r}^2+3x_i^2)-\frac{\xi}{\phi_0}\right]\delta_{ij}.
\end{align}
To express the solution in isotropic coordinates, the following transformation is used \cite{ozer2018linearized,bernabeu2010cosmological}
\begin{equation}
\bar{x}^{i}=x^{i}+\frac{V_0}{24\phi_0}{x^{i}}^3,
\end{equation}
and, then we will get the following solutions 
\begin{align}
h_{00} &=-\left(\frac{2M}{\phi_0 r}+\frac{V_0}{6}r^2+\frac{\xi}{\phi_0}\right),\\
h_{ij} &=-\left(\frac{2M}{\phi_0 r}-\frac{V_0}{12}r^2-\frac{\xi}{\phi_0}\right)\delta_{ij},\\
\xi &=\frac{2M}{(2\omega+3)}\frac{\exp(-m_s r)}{r}-\frac{V_0}{3(2\omega+3)}r^2.
\end{align}
The full metric components will be given by
\begin{equation}
g_{00}= 1-\frac{2M}{\phi_0 r}\left(1+\frac{\exp(-m_s r)}{2\omega+3}\right)- \frac{V_0 r^2}{6\phi_0}\left(1-\frac{2}{2\omega+3}\right),
\end{equation}
\begin{equation}
g_{ij}= -\left[1+\frac{2M}{\phi_0 r}\left(1-\frac{\exp(-m_s r)}{2\omega+3}\right) -\frac{V_0 r^2}{12\phi_0}\left(1-\frac{4}{2\omega+3}\right)\right]\delta_{ij},
\end{equation}
while the scalar field takes the form
\begin{equation}
\phi=  \phi_0\left(1+\frac{2M}{(2\omega+3)}\frac{\exp(-m_s r)}{\phi_0 r} -\frac{V_0}{3\phi_0(2\omega+3)}r^2\right).
\end{equation}
Let us define the following symbols similarly to the previous sections 
\begin{equation}\label{40}
\frac{2\alpha}{c^2}\equiv\frac{2M}{\phi_0 r c^2}\left(1+\frac{\exp(-m_s r)}{2\omega+3}\right)+\frac{V_0 r^2}{6\phi_0}\left(1-\frac{2}{2\omega+3}\right),
\end{equation}
\begin{equation}
\frac{2\beta}{c^2}\equiv \frac{2M}{\phi_0 r c^2}\left(1-\frac{\exp(-m_s r)}{2\omega+3}\right)-\frac{V_0 r^2}{12\phi_0}\left(1-\frac{4}{2\omega+3}\right),
\end{equation}
where, $c^2$, is for dimensional consistency. The expression \eqref{40} reduces to
\begin{equation}\label{41}
\alpha\equiv\frac{M}{\phi_0 r }\left(1+\frac{\exp(-m_s r)}{2\omega+3}\right)+\frac{V_0 r^2}{12\phi_0}\left(1-\frac{2}{2\omega+3}\right).
\end{equation}
Performing the same explicit steps done before for the metric $f(R)$ case, we end up with the following expression for force similar to \eqref{2.128} with the newly defined $\alpha$ \eqref{41} i.e.
\begin{equation}\label{2.179}
\vec{F}=-m\left(-\vec{\nabla}\alpha-\frac{2}{c}\frac{\partial\vec{A}}{\partial t}\right)-\frac{2m}{c}\vec{v}\times (\vec{\nabla}\times\vec{A}).
\end{equation}
For calculating the above let us look at how $\vec{\nabla}\alpha$ looks like
\begin{equation}
\vec{\nabla}\alpha=  -\frac{M}{\phi_0}\frac{1}{r^2}\vec{\hat{r}}-\frac{1}{(2\omega+3)}\frac{M}{\phi_0 r}m_s\exp(-m_s r)\vec{\hat{r}}- \frac{M}{\phi_0}\frac{1}{r^2}\frac{\exp(-m_s r)}{2\omega+3}\vec{\hat{r}} +\frac{V_0}{6\phi_0}r\left(1-\frac{2}{2\omega+3}\right)\vec{\hat{r}}.
\end{equation}
where $\vec{\hat{r}}$ is the radial unit vector. Hence, the expression for force in the generalised Brans-Dicke theory \eqref{2.179} for the stationary case i.e, $\dfrac{\partial \vec{A}}{\partial t}=0$ becomes
\begin{equation}\label{2.181}
\begin{aligned}
\vec{F}_{BD}= & -\frac{Mm}{\phi_0}\frac{1}{r^2}\vec{\hat{r}}-\frac{1}{(2\omega+3)}\frac{Mm}{\phi_0 r}m_s\exp(-m_s r)\vec{\hat{r}}-\frac{Mm}{\phi_0}\frac{1}{r^2}\frac{\exp(-m_s r)}{2\omega+3}\vec{\hat{r}} \\&+ \frac{V_0 m}{6\phi_0}r\left(1-\frac{2}{2\omega+3}\right)\vec{\hat{r}}-\frac{2m}{c}\vec{v}\times (\vec{\nabla}\times\vec{A}),
\end{aligned}
\end{equation}
where $m$ is the mass of the test particle.\\\\
Note that Palatini $f(R)$, when put in the form of a Brans--Dicke class does not gives a vanishing Kinetic term similar to metric $f(R)$. The kinetic term, indeed has a coupling parameter $\omega=-\dfrac{3}{2}$ \cite{sotiriou2010f}.\\\\
From \eqref{2.181}, we see that $\omega=-\dfrac{3}{2}$ is indeed a singular point for some of the terms, and hence, we can safely conclude that such a Lorentz force law in the case of Palatini $f(R)$ is at best ill defined and surely deserves further investigation. 
\subsection{\textbf{Force for a metric $f(R)$ theory with an arbitrary function}}
As seen from \eqref{2.147}, $f(R)$ can be recast as a general scalar--tensor theory with a vanishing kinetic term i.e., $\omega=0$. Then from \eqref{2.181}, we get
\begin{equation}
\begin{aligned}
\vec{F}_{f(R)}= & -\frac{Mm}{\phi_0}\frac{1}{r^2}\vec{\hat{r}}-\frac{1}{3}\frac{Mm}{\phi_0 r}m_s\exp(-m_s r)\vec{\hat{r}}  - \frac{Mm}{\phi_0}\frac{1}{r^2}\frac{\exp(-m_s r)}{3}\vec{\hat{r}} + \frac{V_0 m}{18\phi_0}r\vec{\hat{r}} \\&-\frac{2m}{c}\vec{v}\times (\vec{\nabla}\times\vec{A}).
\end{aligned}
\end{equation}
Which can be seen as the force law for general metric $f(R)$ theories with $\phi=f'(R)$.\\\\
We conjecture that we can use this expression for not only analytic forms of $f(R)$ but also to other models. Of course, we need to check that such models indeed have the correct Newtonian limit and if they pass the solar system constraints. Such $f(R)$ models have to be first subjected to a Legendre transformation to put them into the correct Brans--Dicke form after which we can explicitly calculate the $V_0$ potential.\\\\
It is interesting to note that, the above expression can, in principle, describe a wide range of $f(R)$ theories and hence can be used as a universal Lorentz force law in the context of $f(R)$ theories subject to the condition that a metric variation is chosen. We also make a point that this is not relevant for metric-affine theories because such theories have torsion and non-metricity \cite{sotiriou2007metric} and generalised Brans--Dicke cannot accommodate such features.
\subsection{\textbf{Force for metric $f(R)$ ($f$ analytic)}}
\label{durga}
Let us evaluate $V_0$, using the assumption that $f(R)$ is an analytic function and can be expanded as 
\begin{equation}
f(R)=a_0+a_1 R+\frac{a_2}{2!}R^2...
\end{equation}
Since, $\chi=R $ and $\phi=f'(\chi)=f'(R)$, we have for the expression of potential
\begin{equation}\label{2.184}
\begin{split}
V(\phi) &=R(\phi)f'(R)-f(R)\\
&=R^{(1)}(1+a_2 R^{(1)})-(R^{(1)}+a_2 {R^{(1)}}^2)\\
&=\frac{1}{2}a_2 {R^{(1)}}^2(\phi),
\end{split}
\end{equation}
where we have introduced only the relevant terms of $f(R)$ \eqref{377}\eqref{378}. Now, from eq.\eqref{1.112} we know that $R^{(1)}=\partial_{\mu}\partial_{\rho}h^{\rho\mu}-\Box h$ which is at least second order in the derivatives of the metric perturbation. Hence, we can neglect it in the expression of force which is linear in the derivatives of the metric perturbation, and hence end up with the following expression for force
\begin{equation}\label{2.185}
\vec{F}_{f(R)}=-\frac{Mm}{\phi_0}\frac{1}{r^2}\vec{\hat{r}}-\frac{1}{3}\frac{Mm}{\phi_0 r}m_s\exp(-m_s r)\vec{\hat{r}}-\frac{Mm}{\phi_0}\frac{1}{r^2}\frac{\exp(-m_s r)}{3}\vec{\hat{r}} -\frac{2m}{c}\vec{v}\times\vec{B}.
\end{equation}
Now let us compare it with the expression of force that we obtained from the linearised theory directly which is given by \eqref{2.131}
\begin{equation}\label{2.186}
\vec{F}_{f(R)}=-m\vec{E}-\frac{2m}{c}\vec{v}\times\vec{B}-m\vec{E}\frac{\exp(-\Upsilon r)}{3} -\frac{m}{3}\phi\Upsilon \exp(-\Upsilon r) \vec{\hat{r}}.
\end{equation}
where $\vec{E}=\dfrac{GM}{r^2}\vec{\hat{r}}$. Immediately, we observe that for the two approaches to be equivalent, we need that $m_s=\Upsilon$ (Keep in mind that in the Lorentz force law for Brans--Dicke with a potential, the gravitational constant, $G$, is taken to be unity).
So, from \eqref{2.162}, we must have the following
\begin{equation}\label{2.187}
\frac{\phi_0}{2\omega+3}\frac{d^2V(\phi_0)}{d\phi^2}=m_s^2=\Upsilon^2.
\end{equation}
From \eqref{2.184} and using \eqref{377}\eqref{378}, i.e., 
\begin{align}\label{379}
f(R) &=R^{(1)}+\frac{a_2}{2!}{R^{(1)}}^2 & f'(R) &=1+a_2 R^{(1)},
\end{align}
we can prove \eqref{2.187} . Indeed, one can see that \ref{mahadev}
\begin{equation}\label{2.194}
m_s^2=\frac{\phi_0}{3}\left(a_2\left(\frac{dR^{(1)}(\phi)}{d\phi}\right)^2+R^{(1)}(\phi)\frac{d^2R^{(1)}(\phi)}{d\phi^2}\right)\bigg\rvert_{\phi_0},
\end{equation}
and given that
\begin{align}
\phi =f'(\chi)=f'(R),
\end{align}
eq. \eqref{378}, implies
\begin{equation}\label{2.196}
R^{(1)}=\frac{\phi-1}{a_2}.
\end{equation}
If we make use of \eqref{2.196} in \eqref{2.194}, the second term in \eqref{2.194} goes to zero and we are left with
\begin{equation}\label{2.197}
m_s^2=\frac{\phi_0}{3a_2}.
\end{equation}
Now, for arguments of stability, $\phi_0$ is the minimum of the potential. So, let us find the minimum of \eqref{2.184}. Let us impose
\begin{equation}
\frac{dV}{d\phi}=a_2 R^{(1)}\frac{dR^{(1)}}{d\phi}\bigg\rvert_{\phi_0}=0.
\end{equation}
Again, from \eqref{2.196}, we have for the above expression
\begin{equation}
a_2\left(\frac{\phi-1}{a_2}\right)\frac{1}{a_2}\bigg\rvert_{\phi_0}=0,
\end{equation}
which finally gives us the minimum value $\phi_0$, i.e., 
\begin{equation}\label{2557}
\phi_0=1.
\end{equation}
Hence, after substituting the above minimum value in \eqref{2.197}, we get the following result
\begin{equation}\label{2.201}
m_s^2=\bigg\lvert\frac{1}{3 a_2}\bigg\rvert.
\end{equation}
We have argued that $a_2$ has to be less than zero for $\Upsilon^2$ to be positive (since $\Upsilon^2=-\dfrac{1}{3a_2}$) as it is required for physical solutions of the massive KG equation for the scalar mode. Also, remember that in this theory, $m_s^2$ is positive as well (see \eqref{2.162}). So, we have to put a minus sign in \eqref{2.201} and hence we get
\begin{equation}
m_s^2=-\frac{1}{3 a_2},
\end{equation}
and we proved equation \eqref{2.187}. This goes on to show that the two approaches, one in which we considered $f(R)$ to be analytic \textit{a priori} and the second one, where we considered a generalised Brans--Dicke theory and searched for an expression of a Lorentz force for metric $f(R)$ is equivalent. This is also a self--consistent non--trivial sanity check for both the approaches. 
\subsection{\textbf{Gravitational time delay in Brans-Dicke with a potential}}
We have from \eqref{2.133} that the gravitational time delay is given by
\begin{equation}\label{1062}
\Delta=\frac{1}{2c}k^{\alpha}k^{\beta}\int_{P_1}^{P_2}h_{\alpha\beta}(x)dl.
\end{equation}
Now, from \eqref{2.152}, we have the following
\begin{equation}\label{1060}
\theta_{\mu\nu}=h_{\mu\nu}-\frac{1}{2}\eta_{\mu\nu}h-\eta_{\mu\nu}\frac{\xi}{\phi_0}.
\end{equation}
The above can be contracted to give
\begin{equation}\label{1059}
\theta=-h-4\frac{\xi}{\phi_0},
\end{equation}
which can than be used to eliminate $h$ in \eqref{1060} to give
\begin{equation}\label{1061}
h_{\mu\nu}=\theta_{\mu\nu}-\frac{1}{2}\eta_{\mu\nu}\theta-\eta_{\mu\nu}\frac{\xi}{\phi_0}.
\end{equation}
When we insert \eqref{1061} in \eqref{1062}, the last two terms above do not contribute to the integral as the propagation vectors are considered null with respect to the flat metric. So, the gravitational time delay in Brans--Dicke is given by
\begin{equation}
\Delta=\frac{1}{2c}\int_{P_1}^{P_2}\theta_{00}dl+\frac{1}{c}\int_{P_1}^{P_2}\theta_{0i}\vec{\hat{k}} dl,
\end{equation}
where $\theta_{0i}$ is the additional term coming from considering a slowly rotating source with angular momentum $\vec{J}$.\\\\
Now, $\theta_{00}$ is given by \eqref{1058} as
\begin{equation}
\theta_{00} =-\frac{4M}{\phi_0}\frac{1}{\bar{r}}+\frac{V_0}{6\phi_0}\bar{r}^2.
\end{equation}
 As usual, we define $\theta_{0i}$ similar to the one in GR as
\begin{equation}
\theta_{0i}\equiv\frac{2A_{i}}{c^2}.
\end{equation}
So, the gravitational time delay in brans--dicke is given by
\begin{equation}
\Delta=\frac{1}{2c^3}\int_{P_1}^{P_2}\Big[-\frac{4M}{\phi_0}\frac{1}{\bar{r}}+\frac{V_0}{6\phi_0}\bar{r}^2\Big]dl+\frac{2}{c^3}\int_{P_1}^{P_2}\vec{A}\cdot\vec{\hat{k}}dl,
\end{equation}
where we have introduced $c$ for dimensional consistency. Hence, the Shapiro time delay is given by
\begin{equation}\label{1021}
\Delta_s=\frac{1}{2c^3}\int_{P_1}^{P_2}\Big[-\frac{4M}{\phi_0}\frac{1}{\bar{r}}+\frac{V_0}{6\phi_0}\bar{r}^2\Big]dl.
\end{equation}
While the gravitomagnetic time delay is given by
\begin{equation}
\Delta_{GM}=\frac{2}{c^3}\int_{P_1}^{P_2}\vec{A}\cdot\vec{\hat{k}}dl.
\end{equation}
or from \cite{ciufolini2003gravitomagnetic}
\begin{equation}
\Delta_{GM}=\frac{2J}{\phi_0 c^4}\left(\frac{1}{r_1}+\frac{1}{r_2}\right)\frac{\vec{\hat{J}}\cdot(\vec{\hat{r_1}}\times\vec{\hat{r_2}})}{1+\vec{\hat{r_1}}\cdot\vec{\hat{r_2}}},
\end{equation}
where $\vec{\hat{r_1}}$ and $\vec{\hat{r_2}}$ are unit vectors indicating the positions of $P_1$ and $P_2$, respectively from the centre of a sphere, and $J$ is the magnitude of the angular momentum of a slowly rotating source and $\vec{\hat{J}}$ is the unit vector along it.
\subsection{\textbf{Time delay in $f(R)$ ($f$ analytic)}}
From eq.\eqref{2.184}, we have that, $V(\phi_0)=\frac{1}{2}a_2 {R^{(1)}}^2(\phi_0)$. However, we have shown that $\phi_0=1$ \eqref{2557}, which is the minimum of the potential. As a result, \eqref{2.196} implies, $R^{(1)}(\phi_0)=0$, and hence $V_0=V(\phi_0)=0$. The second term in \eqref{1021} vanishes and we are left with
\begin{equation}
\Delta_s=\frac{1}{2c^3}\int_{P_1}^{P_2}-\frac{4M}{\phi_0}\frac{1}{\bar{r}}dl.
\end{equation}
As we already saw that $\phi_0=1$, we recover the $f(R)$ result given in \eqref{1022} and \eqref{1023}, which is also the same as the one in GR i.e
\begin{equation}\label{1152}
\Delta_s=-\frac{2GM}{c^3}\int_{P_1}^{P_2}\frac{dl}{r},
\end{equation}
\begin{equation}\label{1153}
\Delta_{GM}=\frac{2GJ}{c^4}\left(\frac{1}{r_1}+\frac{1}{r_2}\right)\frac{\vec{\hat{J}}\cdot(\vec{\hat{r_1}}\times\vec{\hat{r_2}})}{1+\vec{\hat{r_1}}\cdot\vec{\hat{r_2}}},
\end{equation}
where, $G$ is introduced for dimensional consistency.
\subsection{\textbf{Discussion on the time delay in Brans--Dicke with a potential}}
As expected, we indeed see a modification in the shapiro time delay after the introduction of the potential. This provides a finite correction to the time delay result we get from GR. Also, from the Brans--Dicke results of time delay, we can consistently derive our previous results for metric $f(R)$ theory with $f$ analytic, which is a non--trivial sanity check. 
\section{Discussion and outlook}
We have studied a gravitoelectromagnetism formalism for metric $f(R)$ theories. We demonstrated that our result for the expression of Lorentz force in the context of metric $f(R)$ consistently drops down to that of the GR result in the appropriate limit, i.e., when the cut--off frequency of the scalar mode is very high (or alternatively, the Compton wavelength is very low).\\\\
Our result is also in agreement with the fact that $f(R)$ theory is known to have a massless graviton plus a scalar field  \cite{corda2008massive,capozziello2008massive,capozziello2010testing,prasia2014detection}, as we have uncovered only attractive effects from the Lorentz force expression. The expression of the force is non--trivial and interesting because we find that not only the gradient of the potential matters but its absolute value also contributes at very short ranges. The latter is a peculiar feature because not many known forces display this behaviour. We conjecture that this behaviour is an imprint of the higher order corrections of $f(R)$ in the linearised version of the theory. Anyway, this compels us to further investigate this matter.\\\\
We then consider the gravitational time delay. In this case, the extra scalar degree of freedom associated to $f(R)$ does not contribute and we do not get a different result from GR. Of course, it is possible that higher order corrections will break the degeneracy between the two theories. However, we can safely assume that even if higher order corrections do play a role, it would be negligible or highly suppressed.\\\\
We conclude the chapter by presenting our results for the Lorentz force law in the context of Brans--Dicke theory with a potential. The only assumption considered in the theory is that the Brans--Dicke potential is an analytic function. We observe in this case that that $\omega=-3/2$ is a singular point which alludes towards the fact that a Lorentz force in the case of palatini $f(R)$ maybe ill defined, and worth further investigation which, if confirmed, it might lead to another evidence of inconsistency in this class of theories. We demonstrate that the result for Brans--Dicke theory with a potential can be consistently used to derive our previous results for metric $f(R)$ theory, a non--trivial sanity check of the results we discovered for the metric $f(R)$ case.\\\\
Along with that, we also find an expression of Lorentz force law for arbitrary function of $f(R)$. We conjecture that the obtained expression can be used not only for analytic forms of $f(R)$ but also for more general cases. Of course, we need to check that if such a model indeed have the correct Newtonian limit and if they pass the solar system constraints. So, in principle, our expression can describe a wide range of $f(R)$ theories and hence can be used as a universal force law in the context of $f(R)$ theories subject to the condition that a metric variation is chosen.\\\\
Lastly, we consider the gravitational time delay in the case of Brans--Dicke with a potential where we indeed get a correction. However, the correction factors disappears when we reduce it down to that of $f(R)$ with $f$ analytic. \\\\
It is also important to make the point that we did not consider metric--affine theories because such theories have torsion and non--metricity \cite{sotiriou2007metric} and Brans--Dicke cannot accommodate such features.
It would be interesting to see if one could indeed do a GEM formalism for metric--affine theories.\\\\ Anyway, if the latter is done, we can conjecture that the effect of torsion would not be there because macroscopically speaking, in theories with
non propagating torsion we shall need generically spin current fluxes (fermion fluxes) to get locally non--zero torsion but if one assumes the spin to be randomly oriented and not polarized, it will typically average out to zero \cite{sotiriou2007metric}.\\\\ 
It is worthy to mention here that we have explicitly treated the analytic metric $f(R)$ corrections to the Lense--Thirring effect in \cite{dass_liberati}. We have done a post-Newtonian calculation of the gyroscopic frequency in the case of metric $f(R)$ with $f$ analytic and then identified the Lense--Thirring and Geodetic contribution. We find that metric $f(R)$ do not alter the GR Lense--Thirring precession at linear order but it indeed gives a finite correction to the GR Geodetic effect. 
\begin{acknowledgements}
AD would like to thank A. Baldazzi for useful discussions. The authors would like to thank M. Rinaldi for useful feedback. This work was done in SISSA, Italy as part of master's thesis by AD. 
\end{acknowledgements}
\section{\textbf{Appendix: Evaluation of the potential term}}\label{mahadev}
From \eqref{2.184}, we get for the first derivative of $V(\phi)$ as
\begin{equation}
\frac{d V(\phi)}{d\phi}=\frac{d R(\phi)}{d\phi}f'(R(\phi))+R(\phi)\frac{d}{d\phi}f'(R(\phi))-\frac{d}{d\phi}f(R),
\end{equation}
and consequently the second derivative of $V(\phi)$ can be expressed as
\begin{equation}\label{2.189}
\begin{aligned}
\frac{d^2V(\phi)}{d\phi^2} = &\frac{d^2R(\phi)}{d\phi^2}f'(R(\phi))+\frac{dR(\phi)}{d\phi}\frac{d}{d\phi}f'(R(\phi))+ R(\phi)\frac{d^2}{d\phi^2}f'(R(\phi)) +\frac{d}{d\phi}f'(R(\phi))\frac{d R(\phi)}{d\phi}\\&-\frac{d^2}{d\phi^2}f(R(\phi)).
\end{aligned}
\end{equation}
Now, using the following quantities which we encountered in the linearised metric $f(R)$ section, i.e, 
\begin{align}
f(R) &=R^{(1)}+\frac{a_2}{2!}{R^{(1)}}^2, & f'(R) &=1+a_2 R^{(1)},
\end{align}
Where $R^{(1)}$ is the linearised Ricci Scalar. In the susbsequent calculation, we drop the suffix $(1)$ and introduce at the end for sake of convenience. Hence, we get the following expression after evaluating for \eqref{2.189}
\begin{equation}
\begin{aligned}
\frac{d^2V(\phi)}{d\phi^2}=&\frac{d^2R(\phi)}{d\phi^2}(1+a_2 R(\phi))+\frac{dR(\phi)}{d\phi}\frac{d}{d\phi}(1+a_2 R(\phi))+R(\phi)\frac{d^2}{d\phi^2}(1+a_2 R(\phi)) \\&+\frac{d}{d\phi}(1+a_2 R(\phi))\frac{d R(\phi)}{d\phi}-\frac{d^2}{d\phi^2}(R(\phi)+\frac{a_2}{2!}R(\phi)^2),
\end{aligned}
\end{equation}
or
\begin{equation}
\begin{aligned}
\frac{d^2V(\phi)}{d\phi^2}=&\frac{d^2R(\phi)}{d\phi^2}+a_2 R(\phi)\frac{d^2R(\phi)}{d\phi^2}+a_2\left(\frac{dR(\phi)}{d\phi}\right)^2+R(\phi)\frac{d^2R(\phi)}{d\phi^2}+a_2\left(\frac{dR(\phi)}{d\phi}\right)^2- \frac{d^2R(\phi)}{d\phi^2}\\&-a_2\left(\frac{dR(\phi)}{d\phi}\right)^2-a_2 R(\phi)\frac{d^2R(\phi)}{d\phi^2},
\end{aligned}
\end{equation}
and hence, the expression drops to just
\begin{equation}
\frac{d^2V(\phi)}{d\phi^2}=a_2\left(\frac{dR(\phi)}{d\phi}\right)^2+R(\phi)\frac{d^2R(\phi)}{d\phi^2}.
\end{equation}
So, if we use the above result, from \eqref{2.187}, we have the following expression
\begin{equation}
m_s^2=\frac{\phi_0}{3}\left(a_2\left(\frac{dR^{(1)}(\phi)}{d\phi}\right)^2+R^{(1)}(\phi)\frac{d^2R^{(1)}(\phi)}{d\phi^2}\right)\bigg\rvert_{\phi_0}
\end{equation}

\bibliographystyle{spphys}       


\end{document}